\documentclass{article}
\usepackage{braket,bm,latexsym,amsmath,amssymb,amsfonts,fancyhdr,color,graphicx,multirow,slashed,cite}
\usepackage[a4paper,bottom=3cm,top=2.5cm,head=0mm,width=17cm,dvipdfm]{geometry}
\usepackage[usenames,dvipsnames,svgnames,table]{xcolor}
\usepackage[colorlinks=true,
            linkcolor=blue,
            urlcolor=blue,
            citecolor=green,          
						bookmarks=true,
						bookmarksnumbered=true,
						breaklinks=true,
						pdfpagemode=Fullscreen,
						pdfstartview=FitBH]{hyperref}
\usepackage[dotinlabels]{titletoc}
\usepackage{titlesec}
\usepackage{authblk,ulem}

\numberwithin{equation}{section}
\allowdisplaybreaks[4]

\titlelabel{\thetitle.\quad \hspace{-0.8em}}
\titlecontents{section}
              [1.5em]
              {\vspace{4mm} \large \bf}
              {\contentslabel{1em}}
              {\hspace*{-1em}}
              {\titlerule*[.5pc]{.}\contentspage}
\titlecontents{subsection}
              [3.5em]
              {\vspace{2mm}}
              {\contentslabel{1.8em}}
              {\hspace*{.3em}}
              {\titlerule*[.5pc]{.}\contentspage}
\titlecontents{subsubsection}
              [5.5em]
              {\vspace{2mm}}
              {\contentslabel{2.5em}}
              {\hspace*{.3em}}
              {\titlerule*[.5pc]{.}\contentspage}

\newcommand{\titledef}{Formation of Supermassive Black Hole Seeds from Vacuum-Dependent Thermal Dark Matter} 

\hypersetup{ pdfauthor = {Jie Sheng},
	     pdftitle = {\titledef}, 
	     pdfsubject = {}, 
             pdfkeywords = {}, 
	     pdfcreator = {LaTeX with hyperref package},
	     pdfproducer = {dvips + ps2pdf} }

\definecolor{gesfblack}{rgb}{0,0,0}

\definecolor{gesfblue}{rgb}{0.08,0.42,0.76}

\definecolor{gesfgreen}{rgb}{0,1,0}

\definecolor{gesfgrey}{rgb}{0.5,0.5,0.5}

\definecolor{gesflanse}{rgb}{0.00,0.50,0.50}

\definecolor{gesfpurple}{rgb}{0.47,0.19,0.42}

\definecolor{gesfred}{rgb}{1,0,0}

\definecolor{gesfwhite}{rgb}{1,1,1}

\definecolor{gesfyellow}{rgb}{0.7,0.4,0.3}

\newcommand{\gfig}[1]{{\hypersetup{linkcolor=violet}Fig.\,\ref{#1}\hypersetup{linkcolor=blue}}}

\definecolor{Orange}{cmyk}{0,0.61,0.87,0}
\definecolor{JungleGreen}{cmyk}{0.99,0,0.52,0}
\definecolor{OliveGreen}{cmyk}{0.64,0,0.95,0.40}
\definecolor{Brown}{cmyk}{0,0.81,1,0.60}
\definecolor{RoyalBlue}{cmyk}{0.71,0.53,0,0.12}
\definecolor{Gray}{cmyk}{0,0,0,0.40}
\definecolor{LightPink}{cmyk}{0.0,0.25,0,0}
\definecolor{LLightPink}{cmyk}{0.0,0.10,0,0}
\definecolor{LightBlue}{cmyk}{0.25,0,0,0}
\definecolor{LightGray}{cmyk}{0,0,0,0.2}

\graphicspath{{figs/}}

\usepackage{subfigure}
\newcommand{\bee}{\begin{equation}}
\newcommand{\ene}{\end{equation}}
\newcommand{\bea}{\begin{eqnarray}}
\newcommand{\ena}{\end{eqnarray}}

\begin{document}
\fontsize{12pt}{14pt}\selectfont

\title{
       \Large \bf \titledef} 

\author[1,2,3]{{\large Masahiro Kawasaki} \footnote{\href{mailto:masahiro.kawasaki@ipmu.jp}{masahiro.kawasaki@ipmu.jp}}}
\author[1]{{\large Jie Sheng} \footnote{\href{mailto:jie.sheng@ipmu.jp}{jie.sheng@ipmu.jp}}}
\author[1]{{\large Tsutomu T. Yanagida} \footnote{\href{mailto:tsutomu.tyanagida@sjtu.edu.cn}{tsutomu.tyanagida@sjtu.edu.cn}
}}

\affil[1]{Kavli IPMU (WPI), UTIAS, University of Tokyo, Kashiwa, 277-8583, Japan}
\affil[2]{ICRR, University of Tokyo, Kashiwa, 277-8582, Japan}
\affil[3]{Physics Division, Faculty of Science, Kanagawa University, Kanagawa 221-8686, Japan}


\date{}

\maketitle

\vspace{-2mm}
\begin{abstract}
\fontsize{10pt}{12pt}\selectfont

The origin of supermassive black holes, especially those already present at high redshift, remains an open problem. 
We propose that their seeds are primordial black holes formed from rare overdense regions of conventional thermal DM. 
A spectator scalar with two degenerate vacua gives the DM different masses, and hence different freeze-out abundances, in the two vacua. 
Inflationary fluctuations populate the vacuum containing the heavier relic only in exponentially rare patches. 
As the Universe cools, the density contrast of these patches grows sufficiently large for the patches to collapse into PBHs after horizon re-entry. 
We derive the resulting PBH mass function and show that it is sharply concentrated near a lower cutoff while reproducing the mass scale and abundance required for SMBH seeds. 
Since the overdensities originate from a rare, non-Gaussian population rather than from an enhancement of nearly Gaussian primordial curvature perturbations, the usual CMB spectral-distortion constraint does not directly apply. 
The construction also correlates the seed population with the dark-sector spectrum and the inflation scale, providing complementary avenues for testing the scenario.

\end{abstract}

\tableofcontents

\section{Introduction}
\label{sec:introduction}

Supermassive black holes (SMBHs) with masses of
$10^{6}$--$10^{10}\,M_\odot$ are known to reside at the centers of most massive galaxies~\cite{Kormendy:1995er,Magorrian:1997hw,Richstone:1998ky}.
Their origin, however, remains an open problem.
In particular, quasars hosting black holes with masses of order $10^{9}\,M_\odot$ have already been observed at redshifts $z\gtrsim 6$, leaving only a limited time for stellar-remnant black holes to grow to such large masses through accretion and mergers~\cite{Matsuoka:2016pho,Volonteri:2010wz,Woods:2018lty}.
This has motivated scenarios with substantially heavier seed black holes.
Numerical studies suggest that black holes with masses around $10^{4}$--$10^{5}\,M_\odot$ formed at high redshift can provide suitable seeds for the observed SMBHs~\cite{Bean:2002kx,Nakama:2016kfq}.

Primordial black holes (PBHs)~\cite{Hawking:1971ei,Carr:1974nx,Carr:1975qj}, which can form long before the onset of star formation, are therefore attractive candidates for the seeds of SMBHs~\cite{Kawasaki:2012kn}.
A conventional possibility is that large primordial curvature perturbations are generated during inflation and subsequently collapse into PBHs
after horizon re-entry~\cite{Garcia-Bellido:1996mdl,Yokoyama:1995ex,Kawasaki:1997ju,Carr:2016drx}.
For massive PBHs relevant for SMBH seeds,  however, a required enhancement of nearly Gaussian perturbations is strongly constrained by CMB spectral distortions induced by Silk damping~\cite{Hu:1994bz,Chluba:2011hw,Kohri:2014lza}.
This motivates mechanisms in which PBHs originate from rare and highly non-Gaussian overdense regions.

One realization of such a mechanism is provided by inhomogeneous Affleck--Dine baryogenesis~\cite{Nakama:2016kfq,Hasegawa:2017jtk,Hasegawa:2018yuy,Kawasaki:2019iis}.
Inflationary fluctuations of the Affleck--Dine field can generate isolated regions with high baryon densities. 
These regions become overdense after the QCD transition and may subsequently collapse into PBHs.
Such a mechanism can evade the usual CMB spectral distortion constraint, but relies on the specific dynamics of the Affleck--Dine field and is subject to additional cosmological constraints, including those associated with the spatial clustering of rare PBH-forming regions~\cite{Kawasaki:2021zir}.

PBHs can also be generated from inhomogeneities directly in the dark sector~\cite{Xiao:2021ftk,Kawana:2021tde,Flores:2023zpf}.
For example, rare QCD-axion bubbles with anomalously large local axion abundance can collapse into PBHs and provide SMBH seeds~\cite{Ferrer:2018uiu,Kitajima:2020kig,Li:2023zyc,Kasai:2023ofh,Kasai:2023qic}.
More generally, sufficiently large cold dark matter (DM) isocurvature perturbations can also lead to PBH formation~\cite{Passaglia:2021jla,Qin:2025ymc}.
These examples demonstrate that the overdensity responsible for PBH formation need not originate from the curvature perturbations.
Existing realizations, however, typically involve nonthermal DM production or primordial DM isocurvature fluctuations.

In this paper, we consider instead a conventional thermal freeze-out DM scenario in which the DM relic density depends on the local vacuum.
We introduce a real spectator scalar field $\Phi$~\cite{Maeso:2021xvl} with two degenerate vacua and a DM particle $\chi$ whose mass depends on the vacuum expectation value of $\Phi$ through a Yukawa interaction $y_\chi\Phi\bar{\chi}\chi$.
Inflationary fluctuations of $\Phi$ generate rare patches that settle into the second vacuum, where the DM particle is heavier and consequently has a larger thermal relic density.
These heavy-DM bubbles (HDBs) become increasingly overdense relative to radiation background and, if sufficiently large, collapse into PBHs after horizon re-entry. The resulting PBHs can serve as seeds for SMBHs.

The rest of this paper is organized as follows.
In Sec.~\ref{sec:model}, we introduce the physical picture and the vacuum-dependent DM model.
In Sec.~\ref{sec:stochastic-evolution}, we study the stochastic evolution of the spectator field during inflation and derive the distribution of HDBs.
In Sec.~\ref{sec:thermal-relics}, we discuss thermal freeze-out in the two vacua and the condition for PBH formation.
In Sec.~\ref{sec:smbh-parameter-space}, we study the resulting PBH population and the parameter space relevant for SMBH seeds. We also show the phenomenological implications of our scenario.
We conclude in Sec.~\ref{sec:conclusion}.

\section{Physical Picture and Model Setup}
\label{sec:model}

\begin{figure}[t]
\centering
\includegraphics[width=0.80\textwidth]{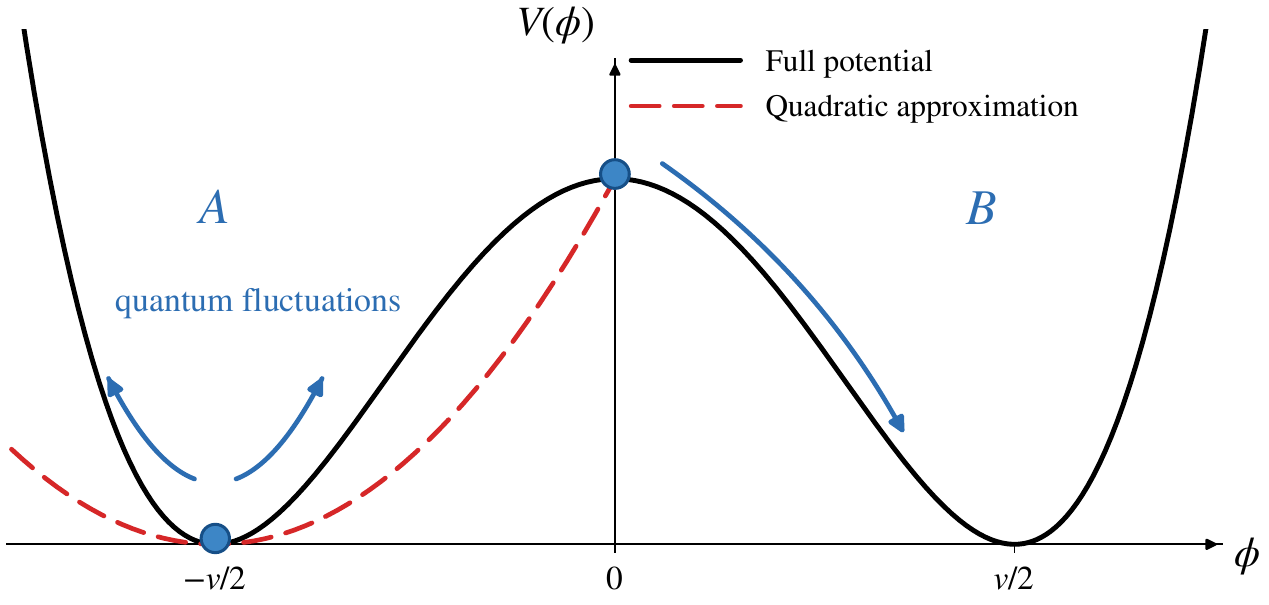}
\caption{Schematic double-well potential for the spectator field.  The observable Universe initially occupies vacuum A at $\Phi=-v/2$.  Inflationary fluctuations generate rare patches that reach the other side of the barrier at $\Phi=0$ and relax to vacuum B at $\Phi=+v/2$.  The dashed curve shows the effective quadratic potential used in the stochastic calculation.}
\label{fig:potential}
\end{figure}

We consider a real spectator field $\Phi$ with a symmetric double-well potential as shown in \gfig{fig:potential},
\begin{equation}
V_0(\Phi)
=
\lambda\left(\Phi^2-\frac{v^2}{4}\right)^2,
\label{eq:double-well-potential}
\end{equation}
where $\lambda>0$ is the quartic coupling and $v$ sets the separation between the two degenerate vacua. 
We refer to the minima at $\Phi=-v/2$ and $\Phi=+v/2$ as vacuum A and vacuum B, respectively. 
The field $\Phi$ is assumed to be energetically subdominant during inflation and therefore does not affect the inflationary background dynamics.

The mass of a stable DM Dirac fermion $\chi$ depends on the local vacuum of $\Phi$.  A minimal interaction realizing this possibility is
\begin{equation}
\mathcal{L}_{\chi}
=
\overline{\chi}i\gamma^\mu\partial_\mu\chi
-\left(m_{\chi}^{(0)}+y_{\chi}\Phi\right)\overline{\chi}\chi
+\mathcal{L}_{\rm ann},
\label{eq:dm-lagrangian}
\end{equation}
where $m_{\chi}^{(0)}$ is a vacuum-independent mass, $y_{\chi}$ is the Yukawa coupling to $\Phi$, and $\mathcal{L}_{\rm ann}$ contains the interactions responsible for thermalizing and depleting $\chi$.  
The physical dark-matter masses in the two vacua are then
\begin{equation}
m_{\chi}
\equiv
m_{\chi}^{(0)}-\frac{y_{\chi}v}{2},
\qquad
m_L
\equiv
m_{\chi}^{(0)}+\frac{y_{\chi}v}{2}.
\label{eq:dm-masses-two-vacua}
\end{equation}
We take $m_L>m_{\chi}$.  
Vacuum A therefore corresponds to the ordinary cosmological region containing the light relic $\chi$, whereas the rare regions that settle into vacuum B contain the heavier state with mass $m_L$. 

The Yukawa interaction in Eq.~\eqref{eq:dm-lagrangian} is the standard
structure of singlet-fermion DM coupled to a real scalar mediator~\cite{Kim:2008pp,Arcadi:2017kky,Arcadi:2019lka}.
Since $\Phi$ is a gauge-singlet scalar, it may communicate with the SM through the Higgs portal $\Phi^2H^\dagger H$. 
After electroweak symmetry breaking, the resulting $\Phi$--Higgs mixing provides a renormalizable realization of fermionic Higgs-portal DM. 
Depending on the mass spectrum and couplings, thermal freeze-out can proceed either through annihilation into SM states mediated by the mixed scalars contained in $\mathcal{L}_{\rm ann}$~\cite{Baek:2011aa,Arcadi:2017kky,Arcadi:2019lka} or, in the secluded regime~\cite{Pospelov:2007mp,Sheng:2026aro}, dominantly through $\chi\bar{\chi}\to\Phi\Phi$.
The cosmological novelty of our setup is that a simple symmetric double-well potential for $\Phi$, with two degenerate vacua, provides a minimal framework for generating PBHs that can serve as SMBH seeds.

Before inflation, different causally disconnected regions may occupy either minimum of Eq.~\eqref{eq:double-well-potential}.  
We assume that the Hubble patch which evolves into our observable Universe initially lies in vacuum A.  
Inflation stretches this patch far beyond its original size, while quantum fluctuations of $\Phi$ occasionally produce exponentially rare subregions where the field fluctuates beyond the barrier and subsequently relaxes into vacuum B.  
Because these regions are rare and do not percolate, the associated domain walls form closed surfaces rather than an extended network.  
We assume that their energy density remains subdominant and that the closed walls collapse after horizon re-entry without erasing the enclosed dark-matter overdensity.  
Under these conditions the usual cosmological domain-wall problem is avoided, since the produced domain walls are isolated and closed surfaces, which collapse after re-entry  of the horizon.

The heavier DM in vacuum B freezes out with a much larger local relic density than that in vacuum A.  
We refer to the resulting rare regions as HDBs.  
At horizon re-entry, if the density contrast of an HDB exceeds the threshold for gravitational collapse, the HDB collapses into a PBH, which can subsequently serve as a seed for an SMBH. 
The remainder of this work quantifies, in sequence, the inflationary production of HDBs, the condition for their collapse, and the abundance of the resulting PBH seeds.

For convenience, we shift the field coordinate so that vacuum A is at the origin:
\begin{equation}
\phi
\equiv
\Phi+\frac{v}{2}.
\label{eq:shifted-field}
\end{equation}
In this coordinate system, vacuum A, the potential barrier, and vacuum B are located at $\phi=0$, $\phi=v/2$, and $\phi=v$, respectively. 
Equation~\eqref{eq:double-well-potential} then becomes
\begin{equation}
V_0(\phi)
=
\lambda\left[\left(\phi-\frac{v}{2}\right)^2-\frac{v^2}{4}\right]^2
=
\lambda\left(v^2\phi^2-2v\phi^3+\phi^4\right).
\label{eq:shifted-potential}
\end{equation}
We denote the threshold separating the two basins of attraction by
\begin{equation}
\phi_c
=
\frac{v}{2}.
\label{eq:critical-field}
\end{equation}
A patch with $\phi>\phi_c$ at the relevant coarse-graining time is assigned to vacuum B, whereas a patch with $\phi<\phi_c$ relaxes back to vacuum A. 
This field redefinition does not affect the calculated probability of crossing the barrier.

To obtain an analytic description of the stochastic evolution, we replace the potential between vacuum A and the barrier by an effective quadratic potential as shown in the red dashed line in \gfig{fig:potential},
\begin{equation}
V_{\rm inf}(\phi)
\simeq
\frac{1}{2}c_I H_I^2\phi^2,
\label{eq:quadratic-potential}
\end{equation}
where $H_I$ is the approximately constant Hubble parameter during inflation and $c_I>0$ is a dimensionless effective mass parameter.  
We normalize this approximation by matching the height of the barrier,
\begin{equation}
V_{\rm inf}(\phi_c)
=
V_0(\phi_c)
=
\frac{\lambda v^4}{16},
\qquad
c_I H_I^2
=
\frac{\lambda v^2}{2}.
\label{eq:barrier-matching}
\end{equation}
We then assume that Hubble patches whose coarse-grained field value exceeds $\phi_c$ during inflation are assigned to vacuum B.
This prescription preserves the suppression associated with the rare transition while allowing a closed analytic treatment.  
In the following, $c_I$ is treated an effective parameter characterizing the inflationary dynamics of $\phi$.

\section{Evolution of the Scalar Field and Distribution of HDBs}
\label{sec:stochastic-evolution}

\subsection{Quantum fluctuations and the Langevin equation}
\label{subsec:langevin}

On super-Hubble scales, the coarse-grained field is overdamped by the inflationary expansion. 
Neglecting $\ddot{\phi}$, its classical equation of motion is
\begin{equation}
3H_I\dot{\phi}
\simeq
-\frac{dV_{\rm inf}}{d\phi}.
\label{eq:slow-roll-field}
\end{equation}
We count the number of e-folds from the beginning of the stochastic evolution and define
\begin{equation}
N
\equiv
\ln\left(\frac{a}{a_{
\rm ini}}\right),
\qquad
dN
=
H_I\,dt,
\label{eq:efold-definition}
\end{equation}
where $a$ is the scale factor and $a_{\rm ini}$ is its value at $N=0$.
Using Eqs.~\eqref{eq:quadratic-potential} and \eqref{eq:slow-roll-field}, the deterministic drift is written as
\begin{equation}
\frac{d\phi}{dN}
\simeq
-\frac{1}{3H_I^2}\frac{dV_{\rm inf}}{d\phi}
=
-\frac{c_I}{3}\phi.
\label{eq:classical-drift}
\end{equation}

For a sufficiently light field, inflation also generates stochastic fluctuations with a characteristic amplitude per e-fold of
\begin{equation}
\delta\phi
\simeq
\frac{H_I}{2\pi}.
\label{eq:quantum-kick}
\end{equation}
The long-wavelength dynamics can therefore be described by the Langevin equation
\begin{equation}
\frac{d\phi}{dN}
=
-\frac{c_I}{3}\phi
+\frac{H_I}{2\pi}\xi(N),
\label{eq:langevin-equation}
\end{equation}
where $\xi(N)$ is Gaussian white noise satisfying
\begin{equation}
\left\langle\xi(N)\right\rangle
=
0,
\qquad
\left\langle\xi(N)\xi(N')\right\rangle
=
\delta(N-N').
\label{eq:white-noise}
\end{equation}
The first term in Eq.~\eqref{eq:langevin-equation} drives the field toward vacuum A, while the second term induces diffusion along the real field axis.  Equation~\eqref{eq:langevin-equation} describes a one-dimensional Ornstein--Uhlenbeck process.

Let $P(N,\phi)$ denote the probability density per unit field value for the coarse-grained field to take the value $\phi$ after $N$ e-folds.  
We impose the localized initial condition $P(0,\phi)=\delta(\phi)$, which is equivalent to setting $\phi(0)=0$ in each realization before the stochastic kicks are applied.  
The solution of Eq.~\eqref{eq:langevin-equation} is
\begin{equation}
\phi(N)
=
\frac{H_I}{2\pi}
\int_0^N dN'\,
\exp\left[-\frac{c_I}{3}(N-N')\right]\xi(N').
\label{eq:langevin-solution}
\end{equation}
Because Eq.~\eqref{eq:langevin-solution} is linear in Gaussian noise, $\phi(N)$ is a Gaussian random variable.  
Its mean vanishes, $\left\langle\phi(N)\right\rangle = 0$, and its variance is, 
\begin{equation}
\sigma^2(N) =
\left(\frac{H_I}{2\pi}\right)^2
\frac{1-\exp(-2c_I N/3)}{2c_I/3}.
\label{eq:field-variance}
\end{equation}
For $c_I N\ll1$, Eq.~\eqref{eq:field-variance} reduces to the massless random-walk result $\sigma^2(N)\simeq[H_I/(2\pi)]^2N$.  
At late times the restoring force balances diffusion and the variance approaches $3H_I^2/(8\pi^2c_I)$.

\subsection{Fokker--Planck distribution and HDB volume fraction}
\label{subsec:fokker-planck}

Equivalently, the probability density $P(N,\phi)$ for an ensemble of Hubble patches obeys the Fokker--Planck equation
\begin{equation}
\frac{\partial P(N,\phi)}{\partial N}
=
\frac{\partial}{\partial\phi}
\left[
\frac{1}{3H_I^2}\frac{dV_{\rm inf}}{d\phi}P(N,\phi)
+\frac{H_I^2}{8\pi^2}
\frac{\partial P(N,\phi)}{\partial\phi}
\right].
\label{eq:fokker-planck-general}
\end{equation}
The first and second terms in the square brackets represent drift and diffusion, respectively.
With the quadratic potential, Eq.~\eqref{eq:fokker-planck-general} becomes
\begin{equation}
\frac{\partial P}{\partial N}
=
\frac{\partial}{\partial\phi}
\left[
\frac{c_I}{3}\phi P
+\frac{H_I^2}{8\pi^2}\frac{\partial P}{\partial\phi}
\right].
\label{eq:fokker-planck-quadratic}
\end{equation}
The solution satisfying the initial condition $P(0,\phi)=\delta(\phi)$ is
\begin{equation}
P(N,\phi)
=
\frac{1}{\sqrt{2\pi\sigma^2(N)}}
\exp\left[-\frac{\phi^2}{2\sigma^2(N)}\right]
\label{eq:probability-distribution}
\end{equation}
where $\sigma^2(N)$ is given by Eq.~\eqref{eq:field-variance}.
The distribution is normalized as
\begin{equation}
\int_{-\infty}^{\infty} d\phi\,P(N,\phi)=1.
\end{equation}

Only the positive tail $\phi>\phi_c$ lies in the basin of vacuum B.  
The negative tail, although equally probable in the quadratic approximation, does not cross the physical barrier separating the two vacua.  
The fraction of the physical volume that will become HDBs is therefore
\begin{equation}
f_{\rm HDB}(N)
\equiv
\int_{\phi_c}^{\infty}d\phi\,P(N,\phi)
=
\frac{1}{2}
\operatorname{erfc}\left[
\frac{\phi_c}{\sqrt{2}\,\sigma(N)}
\right],
\label{eq:hdb-volume-fraction}
\end{equation}
where $\operatorname{erfc}(z)$ is the complementary error function.  
The HDBs are thus generated by the exponentially suppressed one-sided tail of the field distribution.

Let $V(N)$ denote the physical volume of the inflating region under consideration.  
For approximately constant $H_I$,
\begin{equation}
V(N)
\simeq
H_I^{-3}\exp(3N).
\label{eq:inflating-volume}
\end{equation}
The physical volume that eventually settles into vacuum B is
\begin{equation}
V_{\rm HDB}(N)
=
V(N)f_{\rm HDB}(N).
\label{eq:hdb-physical-volume}
\end{equation}

\subsection{Production rate and size distribution of HDBs}
\label{subsec:hdb-production}

Differentiating Eq.~\eqref{eq:hdb-physical-volume} with respect to the e-folds number gives
\begin{equation}
\frac{dV_{\rm HDB}}{dN}
=
3V_{\rm HDB}
+V(N)\int_{\phi_c}^{\infty}d\phi\,
\frac{\partial P(N,\phi)}{\partial N}.
\label{eq:hdb-volume-derivative}
\end{equation}
The first term describes the expansion of regions that have already entered the HDB tail.  
The second term describes the increase of newly generated HDB volumes.\footnote{
Strictly speaking, this term describes the net increase in the HDB volume fraction, since stochastic diffusion can also drive the field from $\phi>\phi_c$ back to $\phi<\phi_c$.
For the double-well potential, however, a region that crosses $\phi_c$ subsequently settles into vacuum B, so that the decrease in the HDB volume due to such backward diffusion can be neglected.
In the quadratic approximation, backward diffusion may still occur, which can lead to an underestimate of $\beta_{\rm HDB}$.
In this sense, our estimate of the PBH production rate is conservative.
}

To make this explicit, consider the HDB volume produced between $N$ and $N+dN$ and evaluate its fraction at the end of inflation, denoted by $N_e$.  
The newly produced volume is stretched by $\exp[3(N_e-N)]$, while the total volume grows by the same factor.  
The two factors cancel, giving
\begin{equation}
\beta_{\rm HDB}(N)\,dN
\equiv
\frac{\exp[3(N_e-N)]}{V(N_e)}
V(N)\frac{df_{\rm HDB}}{dN}\,dN
=
\frac{df_{\rm HDB}}{dN}\,dN.
\label{eq:hdb-production-definition}
\end{equation}
Here $\beta_{\rm HDB}(N)$ is a differential volume HDB fraction per e-fold.
Using Eqs.~\eqref{eq:field-variance} and \eqref{eq:hdb-volume-fraction}, one obtains
\begin{equation}
\beta_{\rm HDB}(N)
=
\frac{2\sqrt{\pi}}{3\sqrt{3}}
c_I^{3/2}
\frac{\phi_c}{H_I}
\frac{\exp(-2c_I N/3)}
{\left[1-\exp(-2c_I N/3)\right]^{3/2}}
\exp\left[
-\frac{4\pi^2c_I\phi_c^2}
{3H_I^2\left[1-\exp(-2c_I N/3)\right]}
\right].
\label{eq:hdb-production-rate}
\end{equation}
An HDB generated at smaller $N$ corresponds to a larger physical scale.

\subsection{Horizon mass at re-entry}
\label{subsec:horizon-mass}

A comoving mode with wavenumber $k$ exits the Hubble horizon during inflation when $k=aH_I$.  Let $k_*$ be the CMB pivot scale and $N_{\rm CMB}$ be the e-fold at which it exits. 
Since $a\propto\exp(N)$ during inflation, the wavenumber of a mode that exits the horizon at e-fold $N$ is given by
\begin{equation}
k(N)
=
k_*\exp(N-N_{\rm CMB}).
\label{eq:k-of-N}
\end{equation}
Thus, modes leaving the horizon at larger $N$ have larger wavenumbers $k$ and smaller comoving scales.

After inflation, the scales relevant for PBH formation re-enter the horizon during radiation domination. 
The mass enclosed within a Hubble volume at horizon re-entry is
\begin{equation}
M_H
\equiv
\frac{4\pi}{3}\rho H^{-3}
=
4\pi M_{\rm Pl}^2 H^{-1},
\label{eq:horizon-mass}
\end{equation}
where $M_{\rm Pl}$ is the reduced Planck mass and we have used the Friedmann equation $\rho=3M_{\rm Pl}^2H^2$. 
During radiation domination, the Hubble rate is determined by the radiation energy density. 
Moreover, in the absence of significant entropy production, the comoving entropy is conserved, $s a^3={\rm const.}$, with $s\propto g_{*s}T^3$. Normalizing the scale factor to $a_0=1$, these relations can be written as
\begin{equation}
H
=
\frac{\pi}{\sqrt{90}}\,
g_*^{1/2}(T)\,
\frac{T^2}{M_{\rm Pl}},
\qquad
a(T)
=
\frac{T_0}{T}
\left[
\frac{g_{*s,0}}{g_{*s}(T)}
\right]^{1/3},
\label{eq:H-a-radiation}
\end{equation}
where $T_0$ is the present photon temperature, while $g_*(T)$ and $g_{*s}(T)$ denote the effective numbers of relativistic degrees of freedom for the energy density and entropy density.

A mode with comoving wavenumber $k$ re-enters the horizon when $k=aH$. Substituting Eq.~\eqref{eq:H-a-radiation} into this condition gives
\begin{equation}
k
=
\frac{\pi}{\sqrt{90}}\,
\frac{T_0 T}{M_{\rm Pl}}\,
g_*^{1/2}(T)
\left[
\frac{g_{*s,0}}{g_{*s}(T)}
\right]^{1/3}.
\label{eq:k-reentry-temperature}
\end{equation}
In the following, we take $g_{*s}\simeq g_*$, which is a good approximation over the temperature range of interest.
Thus, for approximately constant relativistic degrees of freedom, the re-entry temperature scales as $T\propto k\,g_*^{-1/6}$. 
Since Eq.~\eqref{eq:horizon-mass} gives $M_H\propto H^{-1}\propto g_*^{-1/2}T^{-2}$, one then obtains $M_H\propto g_*^{-1/6}k^{-2}$. 
Using $T_0=2.725\,{\rm K}$, $g_{*s,0}\simeq 3.91$, and the standard unit conversions fixes the overall normalization. 
The inflationary scale that exits the horizon at e-fold $N$ is related to the CMB pivot scale by Eq.~\eqref{eq:k-of-N}. 
We therefore obtain
\begin{equation}
M_H(N)
\simeq
19.3\,M_\odot
\left(
\frac{g_*}{10.75}
\right)^{-1/6}
\left[
\frac{
k_*\exp(N-N_{\rm CMB})
}{
10^6\,{\rm Mpc}^{-1}
}
\right]^{-2}.
\label{eq:horizon-mass-N}
\end{equation}
Here $g_*=10.75$ is used as a convenient reference value appropriate to the MeV epoch relevant for $M_H\sim 10^4$--$10^5\,M_\odot$. 
With the CMB pivot scale $k_*=0.002\,{\rm Mpc}^{-1}$, $N_e=60$, and $N_{\rm CMB}=10$, Eq.~\eqref{eq:horizon-mass-N} can be inverted as
\begin{equation}
N(M_H)
\simeq
-\frac{1}{2}\ln\left(\frac{M_H}{M_\odot}\right)
+21.5+N_{\rm CMB} = 31.5
-\frac{1}{2}\ln\left(\frac{M_H}{M_\odot}\right).
\label{eq:N-of-horizon-mass}
\end{equation}

The same radiation-era normalization gives the relation between horizon mass and re-entry temperature. 
Substituting Eq.~\eqref{eq:H-a-radiation} into Eq.~\eqref{eq:horizon-mass} gives
\begin{equation}
T_{\rm re}(M_H)
\simeq
434\,{\rm MeV} 
\left(\frac{M_H}{M_\odot}\right)^{-1/2}\,, \quad  M_H
\simeq
\left(\frac{434\,{\rm MeV}}{T_{\rm re}}\right)^2M_\odot
\label{eq:temperature-mass-relation}
\end{equation}
The large HDBs relevant for SMBH seeds re-enter the horizon relatively late, at temperatures below the MeV scale.

\section{Thermal Dark-Matter Relics and PBH Formation}
\label{sec:thermal-relics}

\subsection{Relic-density contrast between the two vacua}
\label{subsec:relic-contrast}

We assume that $\chi$ reaches thermal equilibrium and undergoes standard nonrelativistic freeze-out in both types of regions.  
The two local cosmological histories have the same interaction structure but different masses, $m_\chi$ in vacuum A and $m_L$ in vacuum B.
For either history, labelled by $i=\chi,L$, the asymptotic yield is parametrically given by~\cite{Lee:1977ua,Kolb:1990vq}
\begin{equation}
Y_i^{\infty}
\equiv
\frac{n_i}{s}
\sim
\frac{x_{f,i}}
{M_{\rm Pl}m_i\langle\sigma v\rangle_i},
\qquad
x_{f,i}
\equiv
\frac{m_i}{T_{f,i}},
\label{eq:freezeout-yield}
\end{equation}
where $n_i$ is the number density, $s$ is the entropy density, $T_{f,i}$ is the freeze-out temperature, and $\langle\sigma v\rangle_i$ is the thermally averaged annihilation cross section.  
Since $x_{f,i}$ depends only logarithmically on the microscopic parameters, it is typically of order $20$--$30$ in both vacua.  
After freeze-out, the energy density is $\rho_i=m_iY_i^\infty s$, and hence
\begin{equation}
\frac{\rho_L}{\rho_\chi}
\simeq
\frac{x_{f,L}}{x_{f,\chi}}
\frac{\langle\sigma v\rangle_\chi}
{\langle\sigma v\rangle_L}.
\label{eq:general-density-ratio}
\end{equation}

If the annihilation cross section is controlled by a common dimensionless coupling $g_{\rm ann}$ and no additional mass scale differs between the two vacua, dimensional analysis gives
\begin{equation}
\langle\sigma v\rangle_i
\propto
\frac{g_{\rm ann}^4}{m_i^2}.
\label{eq:annihilation-scaling}
\end{equation}
Neglecting the small difference between $x_{f,L}$ and $x_{f,\chi}$, the ratio of the local relic densities between the two vacua is therefore
\begin{equation}
r
\equiv
\frac{\rho_L}{\rho_\chi}
\simeq
\left(\frac{m_L}{m_\chi}\right)^2.
\label{eq:density-enhancement}
\end{equation}
Equation~\eqref{eq:density-enhancement} is the central particle-physics input of the mechanism.  
This relation assumes ordinary $s$-wave scaling away from thresholds, resonances, and coannihilation regions.
More complicated freeze-out dynamics can be incorporated by using the general expression in Eq.~\eqref{eq:general-density-ratio}.

Because HDBs occupy only a tiny fraction of the total volume, the observed dark-matter abundance fixes the abundance in vacuum A to excellent accuracy.  
In terms of the dimensionless yield $Y_\chi=n_\chi/s$, the measured abundance is conveniently written as
\begin{equation}
m_\chi Y_\chi
=
\frac{\rho_{\chi}}{s}
\simeq
0.44\,{\rm eV}.
\label{eq:observed-dm-yield}
\end{equation}

\subsection{Density contrast and collapse threshold}
\label{subsec:collapse-threshold}

After both relics become nonrelativistic, the total energy densities outside and inside an HDB are
\begin{equation}
\rho_{\rm out}
=
\rho_{\rm rad}+\rho_\chi,
\qquad
\rho_{\rm in}
=
\rho_{\rm rad}+\rho_L,
\label{eq:inside-outside-densities}
\end{equation}
where the radiation density is taken to be initially common to the two regions.  
The density contrast is therefore
\begin{equation}
\delta(T)
\equiv
\frac{\rho_{\rm in}-\rho_{\rm out}}{\rho_{\rm out}}
=
\frac{(r-1)\rho_\chi}{\rho_{\rm rad}+\rho_\chi}
\simeq
(r-1)\frac{\rho_\chi}{\rho_{\rm rad}},
\label{eq:density-contrast-exact}
\end{equation}
where the last approximation applies during radiation domination.  
Using
\begin{equation}
s(T)
=
\frac{2\pi^2}{45}g_{*s}(T)T^3,
\qquad
\rho_{\rm rad}(T)
=
\frac{\pi^2}{30}g_*(T)T^4,
\label{eq:entropy-radiation-densities}
\end{equation}
together with Eq.~\eqref{eq:observed-dm-yield}, we find
\begin{equation}
\delta(T)
\simeq
\frac{4}{3}
\frac{g_{*s}(T)}{g_*(T)}
(r-1)
\frac{0.44\,{\rm eV}}{T}.
\label{eq:density-contrast-temperature}
\end{equation}
The density contrast grows as $T^{-1}$ because the energy density of nonrelativistic DM decreases more slowly than that of radiation.

For a top-hat overdensity entering the horizon during radiation domination, we adopt the approximate collapse criterion $\delta\gtrsim w$, where $w=p_{\rm in}/\rho_{\rm in}$ is the local equation-of-state parameter.  
Since the pressure is dominated predominantly by radiation,
\begin{equation}
w
\simeq
\frac{1}{3(1+\delta)}.
\label{eq:local-equation-of-state}
\end{equation}
Setting $\delta=w$ gives~\cite{Carr:1974nx}
\begin{equation}
\delta_{\rm th}
=
\frac{\sqrt{21}-3}{6}
\simeq
0.26,
\label{eq:collapse-threshold}
\end{equation}
which we use as the PBH-formation threshold.  

Combining Eqs.~\eqref{eq:density-contrast-temperature} and \eqref{eq:collapse-threshold}, an HDB collapses if it re-enters below the critical temperature
\begin{equation}
T_{\rm re} < T_c(r)
\simeq
2.26\,{\rm MeV}
\left(\frac{g_{*s}}{g_*}\right)
\left(\frac{r-1}{10^6}\right).
\label{eq:critical-temperature}
\end{equation}
At MeV temperatures, $g_{*s}\simeq g_*\simeq10.75$, the parameter range of interest roughly requires $r\simeq 10^6 \gg 1$.
Because $M_H\propto T_{\rm re}^{-2}$, the upper bound on the re-entry
temperature translates into a lower bound on the PBH mass,
\begin{equation}
M_{\rm PBH}
\gtrsim
M_c(r)
\simeq
3.7\times10^4\,M_\odot
\left(\frac{10^6}{r}\right)^2.
\label{eq:critical-pbh-mass}
\end{equation}
Here, we identify the PBH mass with the horizon mass at re-entry, $M_{\rm PBH}\simeq M_H$, and set the order-unity collapse efficiency to unity.  
The important consequence is the resulting sharp lower cutoff $M_c$ in the PBH mass spectrum as shown later.

\section{Parameter Space for Seeding SMBHs}
\label{sec:smbh-parameter-space}

\subsection{Abundance of the PBH seeds}
\label{subsec:seed-mass-range}

Observations of quasi-stellar objects (QSOs) at $z\sim 6$~\cite{Willott:2010yu} constrain the SMBH mass function over the range $M_{\rm SMBH}\sim10^8\,M_\odot$--$3\times10^9\,M_\odot$.
Extrapolating the inferred mass function down to $M_{\rm SMBH}=10^6\,M_\odot$, the comoving SMBH mass density is estimated to be
\begin{equation}
\rho_{\rm SMBH}
\simeq
96\,M_\odot\,{\rm Mpc}^{-3}.
\label{eq:smbh-mass-density-z6}
\end{equation}
This corresponds to an SMBH mass fraction relative to DM of 
\begin{equation}
f_{\rm SMBH}
\equiv
\frac{\rho_{\rm SMBH}}{\rho_{\rm DM}}
\simeq
3\times10^{-9}.
\label{eq:smbh-dm-fraction-z6}
\end{equation}
After formation, seed PBHs grow through accretion and mergers; in particular, accretion increases the total mass density in black holes.
Because of the large uncertainty in the subsequent growth history, we estimate the initial seed-PBH mass fraction relative to DM to lie in the range
\begin{equation}
3\times10^{-11}
\lesssim
f_{\rm seed}
\lesssim
3\times10^{-9}.
\label{eq:seed-pbh-fraction}
\end{equation}

Numerical simulations~\cite{Zhu:2020xni} have shown that heavy black-hole seeds with masses of order $10^4$--$10^5\,M_\odot$ can grow into $10^8$--$10^9\,M_\odot$ SMBHs by $z\sim6$--$7$ through efficient accretion and mergers.
Motivated by these results, we consider $M_c=10^4\,M_\odot$ and $M_c=10^5\,M_\odot$ as benchmark lower bounds on the seed PBH mass.
These choices correspond to
\begin{equation}
r =
\begin{cases}
1.9\times10^6 & (M_c=10^4\,M_\odot), \\
6.1\times10^5 & (M_c=10^5\,M_\odot).
\end{cases}
\label{eq:r-benchmark-values}
\end{equation}
Using Eq.~\eqref{eq:density-enhancement}, the corresponding DM mass hierarchies are
\begin{equation}
\frac{m_L}{m_\chi} =
\begin{cases}
1.4\times10^3 & (M_c=10^4\,M_\odot), \\
7.8\times10^2 & (M_c=10^5\,M_\odot).
\end{cases}
\label{eq:dm-mass-ratio-range}
\end{equation}

\subsection{PBH mass function and present abundance}
\label{subsec:pbh-mass-function}

The quantity $\beta_{\rm HDB}(N)dN$ in Eq.~\eqref{eq:hdb-production-rate} is the HDB volume fraction associated with scales leaving the horizon in an interval $dN$.  
We define $\beta_{\rm PBH}(M)d\ln M$ as the PBH energy fraction at formation in a logarithmic mass interval $d\ln M$. 
With $M\equiv M_{\rm PBH}\simeq M_H$, the change of variables gives
\begin{equation}
\beta_{\rm PBH}(M)
=
\left|\frac{dN}{d\ln M}\right|
\beta_{\rm HDB}\bigl(N(M)\bigr)
\Theta\bigl[M-M_c(r)\bigr].
\label{eq:pbh-mass-function-definition}
\end{equation}
According to Eq.~\eqref{eq:N-of-horizon-mass}, $|d N/ d\ln M| = 1/2$. 
The Heaviside function $\Theta$ imposes the collapse threshold at $M=M_c(r)$.

It is convenient to introduce the dimensionless barrier position
\begin{equation}
x
\equiv
\frac{\phi_c}{H_I}
=
\frac{v}{2H_I}.
\label{eq:x-definition}
\end{equation}
For the barrier-matching prescription of Eq.~\eqref{eq:barrier-matching}, $c_I=2\lambda x^2$.  
Substituting Eqs.~\eqref{eq:hdb-production-rate} and \eqref{eq:N-of-horizon-mass} into Eq.~\eqref{eq:pbh-mass-function-definition}, we obtain
\begin{equation}
\begin{aligned}
\beta_{\rm PBH}(M)
={}&
\frac{\sqrt{\pi}}{3\sqrt{3}}
c_I^{3/2}x
\frac{\exp[-2c_I N(M)/3)]}
{\left[1-\exp[-2c_I N(M)/3]\right]^{3/2}}
\\
&\times
\exp\left[
-\frac{4\pi^2c_Ix^2}
{3\left[1-\exp[-2c_I N(M)/3]\right]}
\right]
\Theta\bigl[M-M_c(r)\bigr].
\end{aligned}
\label{eq:pbh-mass-function}
\end{equation}
For fixed $r$, the cutoff is fixed by Eq.~\eqref{eq:critical-pbh-mass}, while the shape and normalization above the cutoff depend only on $c_I$ and $x$.  
The expression is understood to vanish outside the range $0\leq N(M)\leq N_e$ corresponding to modes that exit the horizon during inflation.

The energy fraction carried by PBHs within a given mass interval when they formed can be written as
\begin{equation}
\beta_{\rm PBH}(M)
\simeq
\left.
\frac{1}{\rho_{\rm rad}}
\frac{d\rho_{\rm PBH}}{d\ln M}
\right|_{\rm form}.
\end{equation}
To connect Eq.~\eqref{eq:pbh-mass-function} to the present abundance, we define the differential PBH density parameter per logarithmic mass interval $\Omega_{\rm PBH}(M)$ as
\begin{equation}
\Omega_{\rm PBH}(M)
\equiv
\frac{1}{\rho_{\rm crit,0}}
\frac{d\rho_{\rm PBH,0}}{d\ln M},
\label{eq:beta-omega-definitions}
\end{equation}
where $\rho_{\rm crit,0}$ and $\rho_{\rm PBH,0}$ denote the present critical density and PBH energy density, respectively.
The PBH energy density scales as $\rho_{\rm PBH}\propto a^{-3}$, whereas the radiation energy density is proportional to $g_*T^4$.
Their ratio therefore grows as $(g_*T^4a^3)^{-1}$.
Using entropy conservation, $g_{*s}T^3a^3={\rm const.}$, together with the approximation $g_*\simeq g_{*s}$, the ratio grows approximately as $T^{-1}$ during radiation domination.
At matter-radiation equality,
\begin{equation}
\left.
\frac{\rho_{\rm PBH}}{\rho_{\rm rad}}
\right|_{\rm eq}
\simeq
\frac{T_{\rm re}(M)}{T_{\rm eq}}
\beta_{\rm PBH}(M),
\label{eq:pbh-fraction-at-equality}
\end{equation}
where $T_{\rm eq}$ is the photon temperature at equality.
Since $\rho_m\simeq\rho_{\rm rad}$ at equality and both PBHs and the total matter density subsequently scale as $a^{-3}$, the present differential abundance $f_\mathrm{PBH}(M)$ is
\begin{equation}
f_\mathrm{PBH}(M)
\equiv 
\frac{\Omega_{\rm PBH}(M)}{\Omega_c}
\simeq
\frac{\Omega_m}{\Omega_c}
\frac{T_{\rm re}(M)}{T_{\rm eq}}
\beta_{\rm PBH}(M),
\label{eq:pbh-abundance-evolution}
\end{equation}
where $\Omega_m$ and $\Omega_c$ are the present total-matter and cold DM density parameters.  
Using Eq.~\eqref{eq:temperature-mass-relation}, this becomes
\begin{equation}
f_\mathrm{PBH}(M)
\simeq
\left(\frac{\beta_{\rm PBH}(M)}{1.6\times10^{-9}}\right)
\left(\frac{\Omega_ch^2}{0.12}\right)^{-1}
\left(\frac{M}{M_\odot}\right)^{-1/2},
\label{eq:present-pbh-abundance}
\end{equation}
where $h$ is the reduced Hubble constant defined through $H_0=100h\,{\rm km}\,{\rm s}^{-1}\,{\rm Mpc}^{-1}$. 

\begin{figure}[t]
\centering
\includegraphics[width=0.80\textwidth]{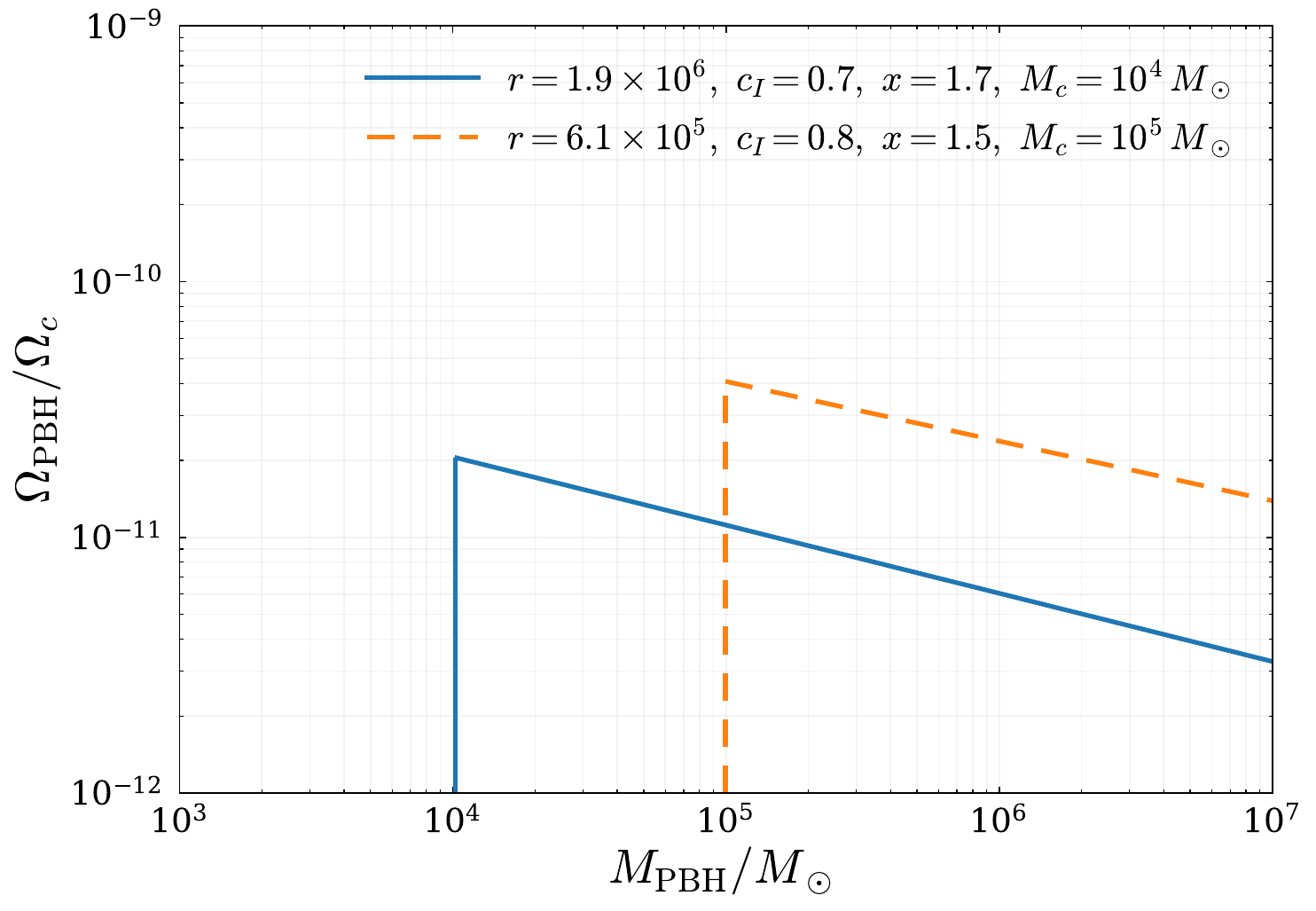}
\caption{Present PBH abundance per logarithmic mass interval for two benchmark parameter choices.
The solid blue curve corresponds to
$(r,c_I,x)=(1.9\times10^{6},\,0.7,\,1.7)$,
for which the collapse threshold is $M_c=10^{4}M_\odot$,
while the dashed orange curve corresponds to
$(r,c_I,x)=(6.1\times10^{5},\,0.8,\,1.5)$,
with $M_c=10^{5}M_\odot$.
The sharp low-mass cutoffs are determined by the PBH collapse condition.}
\label{fig:pbh-abundance}
\end{figure}

Figure~\ref{fig:pbh-abundance} illustrates the characteristic PBH spectrum with certain parameters as explained in the next section.  
The distribution is truncated at $M_c(r)$, producing a sharp peak immediately above the cutoff. 
The resulting population is therefore effectively narrow for SMBH-seeding purposes.
Its total energy fraction is tiny compared with the DM abundance, as required for a rare population of massive seeds.

\subsection{Number density of SMBH seeds}
\label{subsec:smbh-seed-density}

The total mass fraction of PBHs sufficiently massive to serve as SMBH seeds is therefore
\begin{equation}
f_{\rm seed}
=
\int_{\ln M_c}^{\ln M_{\rm max}}
d\ln M\,
f_\mathrm{PBH}(M)
\label{eq:seed-mass-fraction}
\end{equation}
where $M_{\rm max}$ denotes the maximum PBH mass corresponding to HDBs generated at $N=0$. 
The result is insensitive to the precise value of $M_{\rm max}$ because the high-mass tail is exponentially suppressed.
Substituting Eq.~\eqref{eq:N-of-horizon-mass}, Eq.~\eqref{eq:seed-mass-fraction} can be written as the finite integral
\begin{equation}
    f_{\rm seed} (c_I,x;r)= \frac{2}{1.6\times 10^{-9}} \int_0^{N_c}dN\,\beta_{\rm PBH}(N;c_I,x,r)
    \exp\left[-(31.5-N)\right],
    \label{eq:seed-mass-fraction-N-integral}
\end{equation}
where $N_c\equiv N(M_c)$ is the e-fold number corresponding to the cutoff mass $M_c$.
We scan
\begin{equation}
0.2
\leq
c_I
\leq
1,
\qquad
1
\leq
x
\leq
5,
\label{eq:scan-ranges}
\end{equation}
and require the seed-PBH abundance to be consistent with that inferred for the SMBHs at high redshift
\begin{equation}
3\times10^{-11}
\lesssim
f_{\rm seed}
\lesssim
3\times10^{-9}.
\label{eq:seed-density-requirement}
\end{equation}

\begin{figure}[t]
\centering
\includegraphics[width=0.80\textwidth]{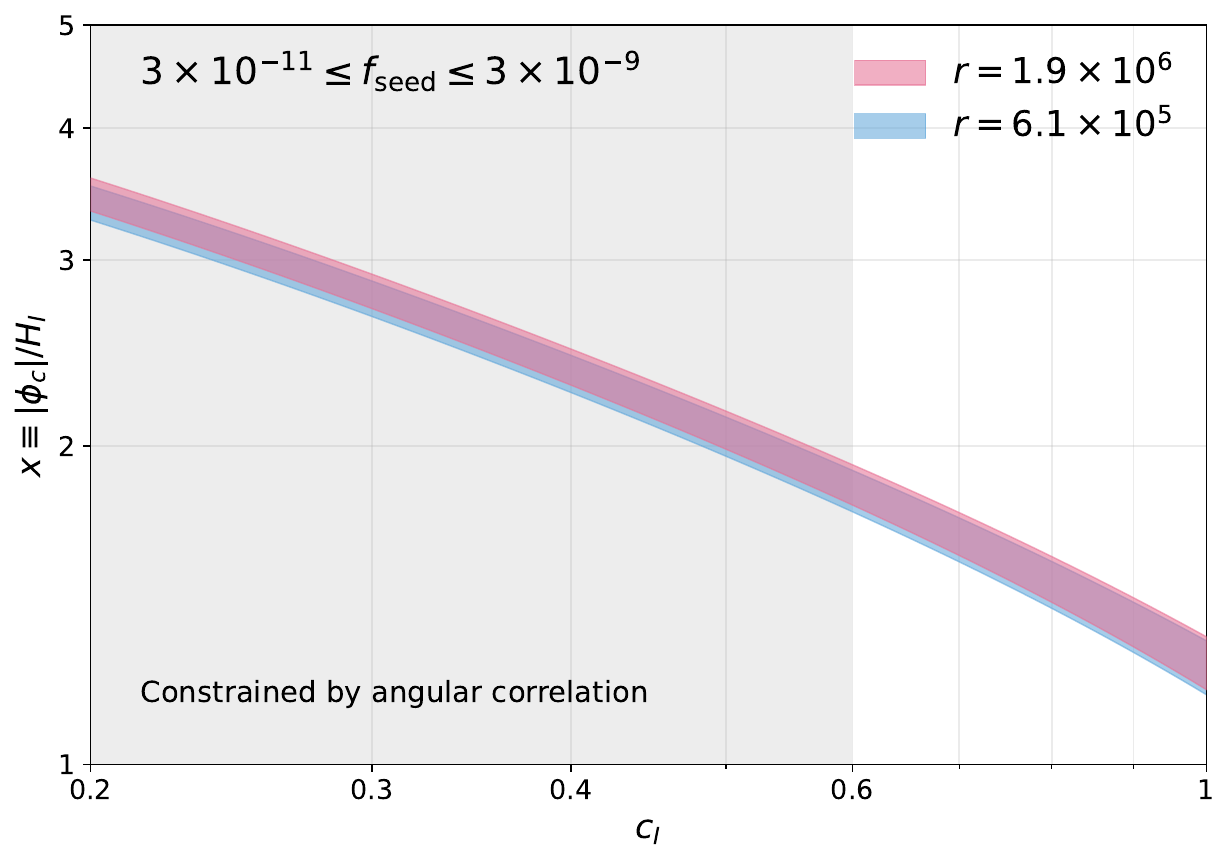}
\caption{
Parameter regions in the $c_I$--$x$ plane that yield
$3\times10^{-11}\lesssim f_{\rm seed}\lesssim3\times10^{-9}$
for $M_c=10^4\,M_\odot$ ($r=1.9\times10^6$) and
$M_c=10^5\,M_\odot$ ($r=6.1\times10^5$).
The shaded region at small $c_I$ is excluded by the angular-correlation constraint from high-redshift quasars.}
\label{fig:parameter-space}
\end{figure}

For $M_c = 10^4\,M_\odot (r=1.9\times 10^6)$ and $M_c=10^5\,M_\odot (6.1\times 10^5)$, the corresponding allowed regions in the $c_I$--$x$ parameter space are shown in \gfig{fig:parameter-space}. 
Increasing $c_I$ strengthens the classical restoring force during inflation and suppresses the probability for the spectator field to fluctuate across the barrier into vacuum B. 
A smaller value of $x=\phi_c/H_I$ is therefore required to maintain the same abundance of SMBH seeds. The dependence on $r$ enters mainly through the lower mass cutoff $M_c(r)$.

An independent constraint arises from the spatial clustering of the seed population.
For small $c_I$, the spectator field is light and its superhorizon fluctuations remain correlated over large scales, causing the rare vacuum-B patches to be strongly clustered.
The resulting angular clustering is constrained by observations of high-redshift quasars, excluding the small-$c_I$ portion of the abundance bands~\cite{Shinohara:2021psq,Shinohara:2023wjd}.
For larger $c_I$, the long-range correlation is more strongly suppressed, and the benchmark value $c_I=0.7$ lies within the allowed region shown in Fig.~\ref{fig:parameter-space}.

\subsection{Implications for the dark-sector parameters and inflation scale}
\label{subsec:inflation-dm-parameters}

The cosmological analysis determines the mass hierarchy between the two vacua, but does not by itself fix the absolute DM mass scale. 
To translate Eq.~\eqref{eq:dm-mass-ratio-range} into particle-physics parameters, we adopt, for orientation, the conventional WIMP-like mass interval for the DM in vacuum A,
\begin{equation}
10\,{\rm GeV}
\lesssim
m_\chi
\lesssim
10^5\,{\rm GeV}.
\label{eq:light-dm-mass-range}
\end{equation}
This interval should be regarded as a representative benchmark rather than a model-independent allowed range. 
The upper endpoint reflects the partial-wave unitarity bound on the annihilation cross section of a pointlike thermal relic, which limits its mass to $\mathcal{O}(10^2)\,{\rm TeV}$~\cite{PhysRevLett.64.615,vonHarling:2014kha}, in the absence of nonstandard annihilation mechanisms~\cite{Harigaya:2016nlg,Kim:2019udq,Kramer:2020sbb,Frumkin:2022ror,Qiu:2023bbp}. 
The lower endpoint is motivated by CMB constraints when freeze-out is controlled by unsuppressed $s$-wave annihilation into electromagnetically interacting final states. 
Such annihilation modifies the ionization history around recombination. 
At fixed relic mass density and annihilation cross section, the injected power increases toward smaller $m_\chi$ because the particle number density is larger~\cite{Kawasaki:2021etm}.
The precise lower limit is therefore channel dependent and can be weakened for velocity-suppressed or sufficiently secluded annihilation~\cite{DAgnolo:2015ujb,Cheng:2022esn,Cheng:2023hzw,Cheng:2023dau}. Combining Eq.~\eqref{eq:light-dm-mass-range} with $m_L/m_\chi=\sqrt{r}$ and the hierarchy in Eq.~\eqref{eq:dm-mass-ratio-range}, we obtain the DM mass range in vacuum B, 
\begin{equation}
7.8\times10^3\,{\rm GeV}
\lesssim
m_L
\lesssim
1.4\times10^8\,{\rm GeV},
\label{eq:heavy-dm-mass-range}
\end{equation}
for $M_c=10^4\,M_\odot$--$10^5\,M_\odot$.
In the parameter region that realizes $m_L\gg m_\chi$, the two contributions to the light-vacuum mass are nearly canceled, $m_\chi^{(0)}\simeq y_\chi v/2$.

The mass of the physical scalar excitation around either minimum of the
potential is
\begin{equation}
m_\Phi^2
=
\left.
\frac{d^2V_0}{d\Phi^2}
\right|_{\Phi=\pm v/2}
=
2\lambda v^2.
\label{eq:physical-scalar-mass}
\end{equation}
Using the barrier-matching condition together with $x=v/(2H_I)$, one obtains
\begin{equation}
\lambda
=
\frac{c_I}{2x^2},
\qquad
m_\Phi
=
\frac{\sqrt{c_I}}{x}\,v
=
2\sqrt{c_I}\,H_I.
\label{eq:scalar-mass-relations}
\end{equation}
Thus, once $x$ and $c_I$ are fixed by the seed abundance, the quartic coupling is no longer independent. 
The two benchmarks in \gfig{fig:pbh-abundance} give $\lambda\simeq 0.12$ and $0.18$ for $(r,x,c_I)=(1.9\times 10^6,1.7,0.7)$ and $(6.1\times 10^5,1.5,0.8)$.
Combining Eq.~\eqref{eq:scalar-mass-relations} with the exact mass splitting gives
\begin{equation}
m_L-m_\chi
=
(\sqrt{r}-1)m_\chi
=
y_\chi v
=
\frac{y_\chi x}{\sqrt{c_I}}\,m_\Phi.
\label{eq:mass_splitting_lab}
\end{equation}
For $\sqrt{r}\gg1$, the scalar mass is therefore related to the DM mass by
\begin{equation}
\frac{m_\Phi}{m_\chi}
\simeq
\begin{cases}
6.8\times10^2/y_\chi,
& (r,x,c_I)=(1.9\times 10^6,1.7,0.7),
\\[2mm]
3.8\times10^2/y_\chi,
& (r,x,c_I)=(6.1\times10^5,1.7,0.7).
\end{cases}
\label{eq:benchmark-scalar-mass}
\end{equation}
For example, $m_\chi=10\,{\rm GeV}$ and $y_\chi=1$ give $m_\Phi\simeq6.8\,{\rm TeV}$ and $3.8\,{\rm TeV}$ for the first and second benchmarks, respectively. 
The lower end of the WIMP mass interval can therefore place the scalar in the multi-TeV regime.
Its collider phenomenology depends on the interaction connecting $\Phi$ to the Standard Model, such as a Higgs-portal coupling, and is not fixed by the cosmological analysis alone.

After electroweak symmetry breaking, the physical scalar can mix with the SM Higgs.  
We denote the mixing angle by $\theta$.  
Exchange of the two
scalar mass eigenstates then induces the spin-independent DM--nucleon cross section
\begin{equation}
\sigma_{\chi N}^{\rm SI}
\simeq
\frac{\mu_{\chi N}^2}{\pi}
\left[
\frac{y_\chi f_Nm_N}{v_h}
\sin\theta\cos\theta
\left(
\frac{1}{m_h^2}
-
\frac{1}{m_\Phi^2}
\right)
\right]^2 ,
\label{eq:DD}
\end{equation}
where $v_h=246\,{\rm GeV}$, $\mu_{\chi N}=m_\chi m_N/(m_\chi+m_N)$, and $f_N \simeq 0.3$ parameterizes the effective scalar Higgs--nucleon matrix element.
For $m_\Phi\gg m_h$, $|\theta|\ll1$, and $m_\chi\simeq40\,{\rm GeV}$, Eq.~(\ref{eq:DD}) becomes numerically
\begin{equation}
\sigma_{\chi N}^{\rm SI}
\simeq
5.6\times10^{-47}\,{\rm cm^2}
\left(
\frac{y_\chi\sin\theta}{10^{-2}}
\right)^2.
\label{eq:DD_numeric}
\end{equation}
The current LZ limit~\cite{LZ:2024sme} around this mass, $\sigma_{\chi N}^{\rm SI}\lesssim2.2\times10^{-48}\,{\rm cm}^2$, therefore corresponds approximately to $|y_\chi\sin\theta|\lesssim2\times10^{-3}$.
A next-generation multi-ton xenon experiment reaching $\sigma_{\chi N}^{\rm SI} \sim 3\times10^{-49}\,{\rm cm}^2$~\cite{Schumann:2015cpa} would probe $|y_\chi\sin\theta| \sim 7\times10^{-4}$.
If both WIMP DM and the scalar particle are discovered in future experiments, the predicted correlation between their masses could provide a test of our model.
As discussed below, an observational determination of the inflationary Hubble scale would provide a complementary probe.

The same relations connect the dark-sector spectrum to the inflationary Hubble scale:
\begin{equation}
H_I
=
\frac{m_L-m_\chi}{2y_\chi x}
=
\frac{\sqrt{r}-1}{2y_\chi x}\,m_\chi
=
\frac{m_\Phi}{2\sqrt{c_I}}.
\label{eq:heavy-mass-inflation-relation}
\end{equation}
For $m_L\gg m_\chi$, this reduces to
\begin{equation}
H_I
\simeq
\frac{m_L}{2y_\chi x}
\sim
\frac{1}{2y_\chi x}
\left(10^4\text{--}10^8\right)\,{\rm GeV}.
\label{eq:inflation-scale-range}
\end{equation}
Since the seed-abundance calculation selects $x=\mathcal{O}(1)$, the value of $H_I$ is controlled mainly by the mass-generating Yukawa coupling. 
For the first benchmark, an order-one Yukawa coupling, $y_\chi=\mathcal{O}(1)$, favors a relatively low inflationary scale.
High-scale inflation can nevertheless be accommodated if the Yukawa coupling is sufficiently suppressed. For example, choosing $m_\chi=10^5\,{\rm GeV}$ together with $y_\chi=10^{-5}$ gives $H_I \simeq 4\times10^{12}\,{\rm GeV}$.
Such a coupling may be insufficient to maintain thermal
equilibrium and control conventional freeze-out by itself.
In this regime, the additional interactions represented by $\mathcal{L}_{\rm ann}$ in Eq.~\eqref{eq:dm-lagrangian} must thermalize and deplete $\chi$, while $y_\chi$ primarily controls the vacuum-dependent mass splitting.

A final comment concerns the possible gravitational-wave (GW) signals associated
with our scenario. 
In principle, GWs can be generated at two distinct stages: during the collapse of an HDB into a PBH and during the subsequent mergers of the resulting seed black holes.
The first signal is associated with the horizon scale at which the HDB collapses. 
Using Eq.~\eqref{eq:horizon-mass}, the corresponding present-day characteristic frequency is
\begin{equation}
f_0(M)
=
\frac{k(M)}{2\pi}
\simeq
6.8\times10^{-11}\,{\rm Hz}
\left(
\frac{M}{10^4M_\odot}
\right)^{-1/2}
\left(
\frac{g_*}{10.75}
\right)^{-1/12}.
\label{eq:formation_frequency}
\end{equation}
This frequency is far below the ranges accessible to terrestrial and space-based interferometers and is also below the frequency range typically
probed by pulsar timing arrays. 
Moreover, an exactly spherical collapse does not emit GWs. 
Introducing an efficiency factor $\epsilon_{\rm GW}$, defined as the fraction of the energy in a PBH-forming region converted into GWs, gives the order-of-magnitude estimate $\Omega_{\rm GW,0}h^2 \simeq 3.5\times10^{-5} \epsilon_{\rm GW}\beta_{\rm PBH}$ with $g_*=10.75$. 
For the benchmark 
$(M_c = 10^5M_\odot, c_I = 0.8, x = 1.5)$,
$\beta_{\rm PBH}(M_c)\simeq 10^{-17}$, and the optimistic choice $\epsilon_{\rm GW}=0.1$, the corresponding present-day amplitude is $\Omega_{\rm GW,0}h^2 \simeq 10^{-22}$.
The GW signal associated with PBH formation is therefore expected to be far
below foreseeable experimental sensitivities.

A potentially stronger signal can arise if the seed black holes subsequently form binaries and merge. For a binary with total mass $M_{\rm tot}$ at redshift $z$, the observed GW frequency near the innermost stable circular orbit is approximately~\cite{Sesana:2004sp}
\begin{equation}
f_{\rm mer,0}
\simeq
4.0\times10^{-3}\,{\rm Hz}
\left(
\frac{M_{\rm tot}}{10^5M_\odot}
\right)^{-1}
\left(
\frac{1+z}{11}
\right)^{-1}.
\label{eq:lisa_isco}
\end{equation}
The $10^4$--$10^7M_\odot$ seed population considered here can therefore produces merger signals in the low-frequency band targeted by LISA. 
Although such mergers are not unique to our mechanism, the observation of a massive seed population with a lower mass cutoff near the predicted $M_c$ would provide a complementary test of the scenario.

\section{Conclusion and Discussions}
\label{sec:conclusion}

We have shown that vacuum-dependent thermal freeze-out of DM provides a simple mechanism for producing PBH seeds of SMBHs. 
The observable Universe predominantly settles into vacuum A, where $\chi$ constitutes the ordinary thermal relic, while inflationary fluctuations generate exponentially rare patches that cross the barrier and settle into vacuum B. 
The larger DM mass in vacuum B leads to a larger thermal relic abundance and hence to the formation of HDBs. 
Their density contrast grows as the Universe cools, and HDBs that re-enter the horizon sufficiently late collapse into PBHs. 
After PBH formation, no macroscopic vacuum-B regions remain and their surrounding
closed domain walls subsequently collapse.
Since these overdensities originate from a rare, highly non-Gaussian population rather than from an enhancement of the nearly Gaussian primordialan curvature power spectrum, the conventional CMB spectral-distortion constraint on supermassive PBH formation does not directly apply.

For the standard $s$-wave freeze-out scaling, the relic-density contrast is approximately the square of mass ratio $r\simeq(m_L/m_\chi)^2$, and the collapse condition introduces a sharp lower cutoff $M_c\simeq3.7\times10^4\,M_\odot(10^6/r)^2$. 
For $M_c=10^4$--$10^5\,M_\odot$, corresponding to $6.1\times10^5\lesssim r\lesssim1.9\times10^6$, the required mass hierarchy is $7.8\times10^2\lesssim m_L/m_\chi\lesssim1.4\times10^3$.
The resulting PBH spectrum is narrowly concentrated immediately above $M_c$.
We find that the benchmark points $r= 1.9\times10^6$ and $6.1\times10^5$ can reproduce the seed-PBH abundance inferred from observations of QSOs at $z \sim 6$ with $0.7 < c_I < 1$ and $1.2 \lesssim x \lesssim 1.8$.
The choice $c_I\gtrsim0.7$ also sufficiently suppresses the long-range correlations constrained by high-redshift QSO observations.

For a representative WIMP-like range $10\,{\rm GeV}\lesssim m_\chi\lesssim10^5\,{\rm GeV}$, the corresponding heavy-state mass is $7.8\times10^3\,{\rm GeV}\lesssim m_L\lesssim1.4\times10^8\,{\rm GeV}$. 
The same parameter relations connect the DM spectrum to the spectator mass and the inflationary scale, allowing complementary tests through DM direct detection, collider searches for the scalar mediator, and a future determination of the inflation scale. 
Gravitational waves emitted directly during HDB collapse are expected around $10^{-11}\,{\rm Hz}$ with an extremely small amplitude, whereas subsequent mergers of the massive seeds may fall within the LISA frequency band.  
Our mechanism therefore provides a simple realization in which conventional thermal DM produces a rare population of massive PBHs with the masses and abundance required to seed the observed SMBHs.

\section*{Acknowledgements}

M. K. is supported by JSPS KAKENHI Grant No. 25K07297. 
J. S. is supported by the Japan Society for the Promotion of Science (JSPS) as a part of the JSPS Postdoctoral Program (Standard) with grant number: P25018. 
T. T. Y. is supported by the Natural Science Foundation of China (NSFC) under Grant No. 12175134, MEXT KAKENHI Grants No. 24H02244.
All authors are supported by the World Premier International Research Center Initiative (WPI), MEXT, Japan (Kavli IPMU).

\providecommand{\href}[2]{#2}\begingroup\raggedright\endgroup

\end{document}